\documentclass[useAMS,usenatbib]{mn2e}

\usepackage{graphicx}

\def\apj{ApJ}
\def\apjl{ApJL}
\def\apjs{ApJS}
\def\mnras{MNRAS}

\def\aap{A\&A}

\def\pasp{PASP}

\newcommand{\gtsima}{$\; \buildrel > \over \sim \;$}
\newcommand{\simgt}{\lower.5ex\hbox{\gtsima}}
\newcommand{\ltsima}{$\; \buildrel < \over \sim \;$}
\newcommand{\simlt}{\lower.5ex\hbox{\ltsima}}

\title[TeV--PeV gamma-ray flare and absorption in Crab Nebula]{Gamma-ray flare and absorption in Crab Nebula: Lovely TeV--PeV astrophysics}
\author[Kohri, Ohira, Ioka]{Kazunori Kohri$^{1,2}$\thanks{E-mail:
kohri@post.kek.jp}, Yutaka Ohira$^{1}$  and Kunihito Ioka$^{1,2}$ \\
$^{1}$Theory Center, Institute of Particle and Nuclear Studies, KEK, 1-1 Oho, Tsukuba
305-0801, Japan \\
$^{2}$The Graduate University for Advanced Studies (Sokendai), 1-1 Oho, Tsukuba
305-0801, Japan
}
\begin{document}

\date{Accepted XXXXX 2012 . Received 10th March 2012; in original form 29th Feburuary 2012 }

\pagerange{\pageref{firstpage}--\pageref{lastpage}} \pubyear{2012}

\maketitle

\label{firstpage}

\begin{abstract}
We spectrally fit the GeV gamma-ray flares recently-observed in
the Crab Nebula by considering a small blob Lorentz-boosted
towards us.
We point out that  the corresponding inverse-Compton flare at TeV--PeV
region is  more enhanced than synchrotron
by a  Lorentz factor square $\sim \Gamma^2$,
which  is already excluding $\Gamma \simgt 200$ and will be detected
by future TeV--PeV observatories, CTA, Tibet  AS + MD  and LHAASO
for $\Gamma \simgt 30$.
We also show that PeV photons emitted from the Crab Nebula are absorbed
by Cosmic Microwave Background radiation through
electron-positron pair creation.
\end{abstract}


\section{Introduction}

It is well-known that the Crab Nebula is one of the brightest objects
in the hard X-ray and gamma-ray sky. Because it  was believed that its
flux is completely steady,  we have used it  as a standard candle to
calibrate  detectors and  instruments in those energy ranges.  Since
the Crab Nebula had already attained a position like a king of strong
and steady sources in the high-energy gamma-ray sky, its impermanence
must be a historic surprise.
 
Quite recently AGILE (Tavani et al. 2011) and Fermi  ( Abdo et
al. 2011; Buehler 2011) reported  day-timescale gamma-ray flares from
the Crab Nebula in ${\cal O}(10^{2})$MeV-- ${\cal O}$(1)GeV region,
which means it is no longer stationary.  According to  the spectral
fitting of the stationary component,  the flares are most likely
produced by synchrotron emission  with an increase in the electron
energy cutoff $E_{{\rm max},e}\sim$ 10 PeV and/or
 in  magnetic field $B\sim$~2~$m$G.
However,  under  the standard particle acceleration,
the synchrotron energy loss limits
the maximum synchrotron photons, irrespective of $B$, below
\begin{equation}
\label{eq:emax}
E_{\rm syn}^{\max} \simeq \frac{9}{4} \frac{m_e c^2}{\alpha}
\simeq 160\ {\rm MeV},
\end{equation}
which is violated in the flares.  Possible solutions include the
relativistic Doppler boost (e.g., Komissarov \& Lyutikov 2011,
Bednarek \& Idec 2011, Yuan et al. 2011), the electric-field
acceleration in the reconnection layer (e.g., Uzdensky et al. 2011),
the sudden concentration of magnetic field (e.g., Bykov et al. 2012),
and a DC electric field parallel to the magnetic field (Sturrock \&
Aschwanden, 2012), but there has been no consensus yet.

In this paper, we consider the relativistic model that a small blob is
Lorentz-boosted towards us, which emits synchrotron radiation beyond
$E_{\rm syn}^{\max}$ (see Buehler et al. 2011 and references therein).
We stress that we can observe the corresponding inverse-Compton flare
which is simultaneously emitted by the same electrons existing in the
boosted blob.  Interestingly, the Lorentz factor $\Gamma$ of the blob
has been already constrained by the current TeV observations (Mariotti
et al. 2010; Ong et al. 2010) and will be further checked by the
future TeV--PeV gamma-ray observations such as CTA, Tibet AS + MD, and
LHAASO, \footnote{See also a similar experiment, HiSCORE (Tluczykont
et al. 2011)} because inverse-Compton emission is more enhanced than
synchrotron by a factor of $\sim \Gamma^2$ approximately.  In
addition, it is remarkable that we must consider an absorption of PeV
photons by Cosmic Microwave Background (CMB) radiation via
electron-positron pair creation even for a Galactic source, which has
not been taken into account so far.  In order to discriminate the
theoretical models and discover the new phenomena of the CMB
absorption, the Crab Nebula is a pretty attractive experimental site
for TeV--PeV astrophysics.

\section{Stationary emission from Crab Nebula}
\label{sec:crab}

\subsection{Theory and Observation}
\label{subsec:data}

First of all, we discuss stationary components of Crab Nebula
emission.  By assuming a broken power-law with an exponential cutoff
for the primary electron spectrum at the emission site, we
parameterize it as
\begin{eqnarray}
    \label{eq:electron}
    \lefteqn{\frac{dn_{e}}{d E_{e}} = } \\
    &&A_{e} E_{e}^{-s_{e}} \left(1+\frac{E_{e}}{E_{\rm
    cb}} \right) ^{-1}
    \left(1+\frac{E_{\rm
    ib}}{E_{e}} \right)^{-1}
    \exp\left(- 
    \frac{E_{e}}{E_{{\rm max},e}} \right), \nonumber 
\end{eqnarray}
with $n_{e}$ number density of electron, $E_{e}$ electron energy,
$E_{{\rm max},e}$ its maximum cutoff energy, $s_{e}$ electron spectral
index, $E_{\rm cb}$ the cooling break energy, $E_{\rm ib}$ the
intrinsic break energy, and $A_{e}$ normalization.  $E_{\rm cb}$ is
determined by equating the age $t_{\rm age}$ with cooling time $t_{\rm
cool}$ due to the synchrotron energy loss.  Here we assume that the
exponent of energy on the exponential shoulder is not two but unity
(Abdo et al. 2010).  The emission below $\sim {\cal O}$(1)~GeV can be
fitted by synchrotron radiation. The observational data were reported
by COMPTEL (Kuiper et al. 2001) and Fermi (Abdo et al. 2010). We adopt
values for parameters, $t_{\rm age}$=1240 yrs, $s_{e}=2.35$, magnetic
field $B=90 \mu$G, the distance to the Earth $d=2.0$~kpc, and the
intrinsic breaking energy $E_{\rm ib}=30$~GeV. For the choice of those
parameters, e.g., see Abdo et al. (2010) and Tanaka \& Takahara
(2010).  For a reference of the distance, see also Trimble
(1973).  Then the cooling energy is $E_{\rm
cb}=1.3$~TeV, and the cutoff energy is fitted to be $E_{{\rm
max},e}=1.5$~PeV.  Note that the corresponding synchrotron cutoff
energy is $\sim 10$ MeV, but the $\nu F_{\nu}$ peak energy is
$\sim$~4--5 times larger than it because of a finite extent of the
distribution.

\begin{figure}
\begin{center}
\includegraphics[width=80mm]{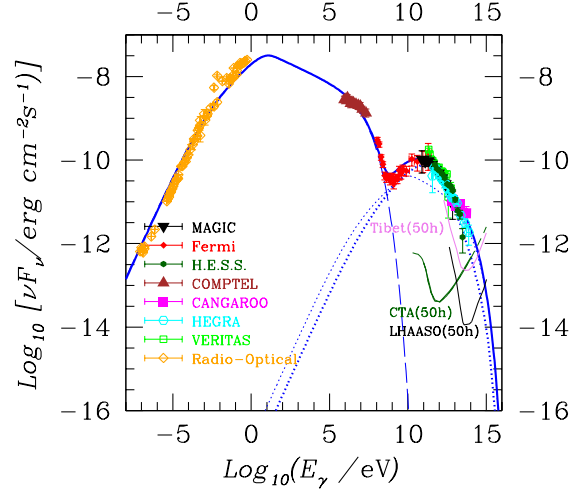}
\end{center}
\caption{Spectrum fitted to  the 
stationary component of radiation from Crab Nebula.  The solid line
shows the total spectrum. The dashed and the dotted lines represent
the synchrotron and the inverse-Compton emissions, respectively. The
upper and the lower curves of  inverse-Compton process at TeV regions
are for scattering off the synchrotron and CMB photons,
respectively. Observational data are plotted as points with their
error bars. We also show sensitivities of future projects such as CTA
(Actis et al. 2011) LHAASO (Cao et al. 2010) and  Tibet AS + MD (Takita
2011) for 50 hours measurements denoted by (50h).}
\label{fig:f1}
\end{figure}

The emission above $\sim {\cal O}(1)$~GeV can be fitted by
inverse-Compton radiation due to the primary electron.  Only the
number density of the CMB photons is too small and insufficient as
target photons to fit the whole data.  Besides we also consider the
synchrotorn photons and adopt the Synchrotron Self-Compton (SSC)
process.  In order to obtain target photon field for the SSC process,
we integrate the photon number density in a volume where the SSC
process occurs.  In a one-zone approximation, we find
\begin{eqnarray}
    \label{eq:SSCtarget}
    \frac{dn_{\gamma,{\rm target}}}{dE_{\gamma}} = 
    \int \frac{dr}{c} \frac{4\pi d^{2}
    }{V }
    \frac{\nu F_{\nu}}{E_{\gamma}^{2}}
    \sim \frac{d^{2} n_{\gamma, {\rm target}}}{dt
    dE_{\gamma} } \frac{R_{\rm SSC}}{c},
\end{eqnarray}
with $E_{\gamma}$ photon energy, $n_{\gamma,{\rm target}}$ the number
density of photon field produced by synchrotron radiation, and $R_{\rm
SSC}$ the effective radius where the SSC process works.  For similar
parameterizations of the effective radius, see Atoyan \& Nahapetian
(1989), Atoyan \& Aharonian (1996), and Tanaka \& Takahara (2010). The
observational data in $\simgt$ TeV regions have been reported by MAGIC
(Albert et al. 2008), HEGRA (Aharonian et al.  2004), CELESTE (Smith
et al. 2006), H.E.S.S. (Aharonian et al. 2006), VERITAS (Celik 2008,
Imran et al. 2009), CANGAROO (Tanimori et al. 1998) in addition to
Fermi (Abdo et al. 2010), Radio and Optical observations (Baars et
al. 1997, Mac{\'{\i}}as-P{\'e}rez et al. 2010). To simultaneously fit
those data, we find the effective radius to be $R_{\rm SSC} \sim
2.0$~pc. In Fig.~\ref{fig:f1} we plot the theoretical fitting and the
observational data. To perform the fitting we use our original code
which has been developed by one of the current authors (KK) in the
series of similar works, e.g., Yamazaki et al. (2006).

\vspace{-0.5cm}

\subsection{PeV gamma-ray absorption by CMB}
\label{subsec:absorption}

\begin{figure}
\begin{center}
\includegraphics[width=80mm]{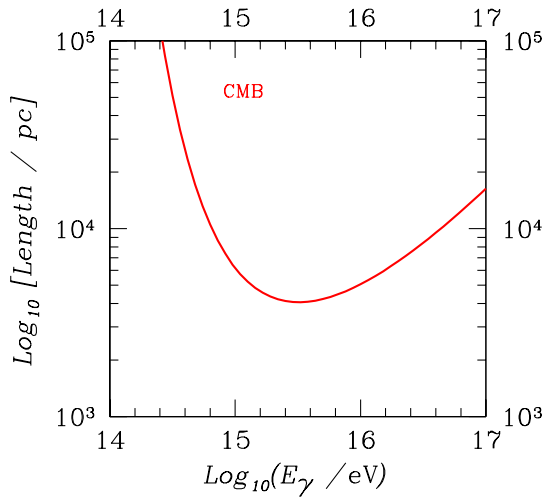}
\end{center}
\caption{Gamma-ray horizon as a function of energy for incident photon. In this
energy region, 
the electron-positron pair production through 
the scattering off
the background CMB
photon dominates the energy-loss rate.}
\label{fig:f2}
\end{figure}

Photon is absorbed if there is a sufficient number of background photons
and the electron-positron pair production  is kinematically allowed
with its energy exceeding threshold $E_{\gamma_{\rm b}} \simgt
m_{e}^{2}c^{4}/(7E_{\gamma})$.  The attenuation length  \footnote{The
attenuation length is equal to the interaction length for our
parameters and the Galactic magnetic field.}   is given by
\begin{eqnarray}
    \label{eq:Latten}
    L_{\rm atten}(E_{\gamma}) = \left[ \int dE_{\gamma_{\rm b}}
    \frac{dn_{\gamma_{\rm b}} }{dE_{\gamma_{\rm b}}} \bar{\sigma} (E_{\gamma},E_{\gamma_{\rm b}}) \right]^{-1},
\end{eqnarray}
where 
\begin{eqnarray}
    \label{eq:barsigma}
    \bar{\sigma} (E_{\gamma},E_{\gamma_{\rm b}})  =
    \int_{-1}^{1-\frac{2m_{e}^{2}c^{4}}{E_{\gamma}E_{\gamma_{\rm b}}}} d \mu
    \frac{1-\mu}{2}
    \sigma_{\rm pair}(E_{\gamma},E_{\gamma_{\rm b}},\mu),
\end{eqnarray}
and the pair-production cross section through $\gamma + \gamma \to
e^{+} + e^{-}$ is given by
\begin{eqnarray}
    \label{eq:sigmapair}
    \lefteqn{\sigma_{\rm pair}(E_{\gamma},E_{\gamma_{\rm b}},\mu) }
    \nonumber \\
    =&& \frac12 \pi
    r_{e}^{2} (1-\beta^{2}) 
    \left[(3-\beta^{4}) \ln\frac{1+\beta}{1-\beta} -2 \beta (2-\beta^{2})\right],
\end{eqnarray}
with $\beta=\sqrt{1-4m_{e}^{2}c^{4} / s}$ and
$s=2E_{\gamma}E_{\gamma_{\rm b}} (1-\mu)$.  When we consider  a
$10^{3}$ TeV ($1$ PeV) photon, the threshold energy of the target
photon for the pair production  becomes the order of ${\cal
O}(10^{-3})$ eV at which the CMB photon dominates along the line of
sight between Crab Nebula and the solar system. In Fig.~\ref{fig:f2}
we plot the attenuation length as a function of the  photon energy
in eV. Remarkably the attenuation length can be down to 4 -- 5 kpc at
$E_{\gamma} \sim$~(2 -- 3)~PeV.

\begin{figure}
\begin{center}
\includegraphics[width=80mm]{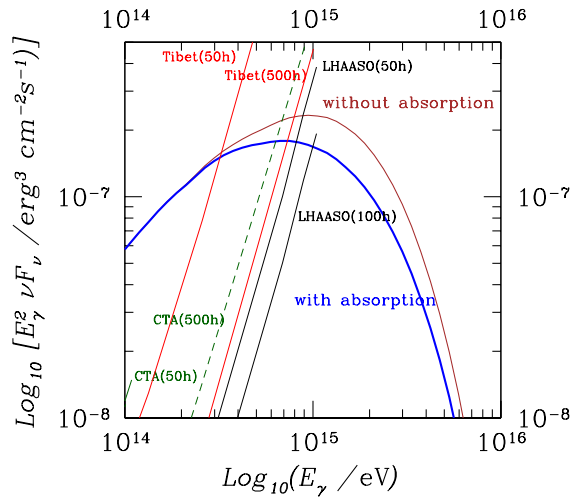}
\end{center}
\caption{Gamma-ray spectra with and without the CMB
absorption. We also plotted sensitivities of future projects  such as
LHAASO (50 h and 100 h), CTA (50 h and 500 h) and Tibet AS + MD  (50 h and 500
h) with their  observation time,
respectively. Note that
the vertical axis means  $\nu F_{\nu}$  times $E_{\gamma}^{2}$.
}
\label{fig:f3}
\end{figure}

\begin{figure}
\begin{center}
\includegraphics[width=80mm]{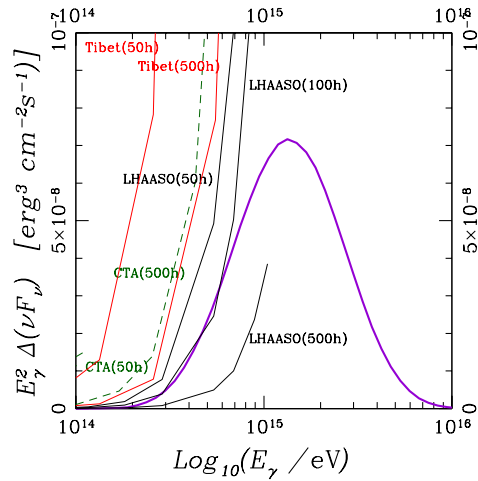}
\end{center}
\caption{Same as Fig.~\ref{fig:f3} but for the difference between two
lines (with and without absorption). Note that the vertical
axis means  $E_{\gamma}^{2}$  times $\Delta(\nu F_{\nu})$ in linear scale.}
\label{fig:f4}
\end{figure}

Because the CMB radiation is ideally homogeneous and isotropic, we can
simply calculate the absorption factor by using $\exp[-d/L_{{\rm
atten}}(E_{\gamma})]$ with $d=2.0$~kpc (Abdo et al. 2010). Multiplying
by this absorption factor, we obtain the observed spectrum.  In
Fig.\ref{fig:f3} we plot spectra with and without this type of
absorption by the CMB radiation. In addition, we also plot
sensitivities of future projects such as LHAASO (50 h and 100 h), CTA
(50 h and 500 h) and Tibet AS + MD (50 h and 500 h) with their
observation time written in round brackets, respectively. In
Fig.~\ref{fig:f4} we compare the observational sensitivities with the
net absorbed component, which is the difference between spectra with
and without the absorption.  From this figure, we find that the
absorption by CMB radiation will be observed by LHAASO with its
observation time of 100 hours.  There could exist
inhomogeneities and uncertainies of the target photons. However, by
calibrating those photons by using more precise observations within 
30 $\%$ at lower eneriges than PeV, we will be able to
distinguish the absorption effect from others.  
In turn,
this means that we have to consider this new type of absorption by the
CMB radiation whenever we observe the PeV photon from Crab Nebula.  As
far as we know, we point out this detectability of the PeV photon
absorption by the future telescopes for the first time.

\section{Fitting to Flare Component}

\begin{figure}
\begin{center}
\includegraphics[width=80mm]{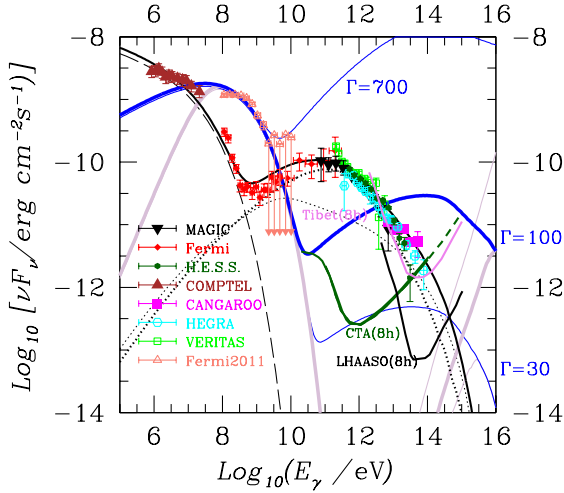}
\end{center}
\caption{Spectrum fitted to the flare component of radiation from
Crab Nebula. We adopted observational data for the flare reported by
Fermi (Buehler 2011, Buehler et al. 2011) which is denoted by
Fermi2011. The thick  solid line shows the theoretical prediction  with
$\Gamma=\delta=10^{2}$, The thin solid lines are the cases for 
 $\Gamma=\delta=30$ and $\Gamma=\delta=700$, respectively. The corresponding models  with
low-energy cutoff ($E_{e,{\rm cut}}=10^{2}$~TeV) are represented by
the shallower solid lines. The meanings of the other lines and the
observational data are the same as those in Fig.~\ref{fig:f1}.}
\label{fig:f5}
\end{figure}

Recently some observations reported that Crab Nebula is no longer
stationary 
 with flares at around ${\cal O}$(1) GeV (Tavani et
al. 2011, Buehler 2011, Buehler et al. 2011). 
Fig.~\ref{fig:f5} shows these data points.
The duration $\Delta t_{\rm obs}$ is
typically the order of one day. In this section we discuss how we can
explain  these flares in terms of synchrotron emission by accelerated
electrons. We consider  Lorentz-boosted blob models in which 
a small blob is boosted with a Lorentz factor $\Gamma$
 and an off-axis viewing angle $\theta$.
In addition, as will be discussed later, it should be natural to
assume accelerated  electrons and  magnetic field in the blob. 
Electrons emit
synchrotron radiation by using the local magnetic field in the
rest  frame of the blob. This model can produce ${\cal O}$(1) GeV
synchrotron photon in the  observer frame unlike the nonrelativistic
models where the synchrotron photon energy cannot exceed   $E_{\rm
syn}^{\max}\sim 10^{2}$ MeV in equation~(\ref{eq:emax}).  $E_{\rm
syn}^{\max}$ is independent of $B$ because $E_{\rm max,e}$ is limited
by balancing synchrotron cooling time with acceleration time.  Here we
do not specify  the origin of the blob,
 which could be the pulsar wind or the
shock at the knot of the inner nebula.

Importantly, inverse-Compton photon  is also emitted simultaneously,
which  is produced by scattering  off both the boosted CMB and
synchrotron radiation  coming from outside the blob.  Both of
synchrotron and inverse-Compton radiation  are boosted-back to
the observer frame, which give larger energy and sizably-larger
fluxes even if the size of the blob is tiny.

Concretely,  the photon energy flux $\nu F_{\nu}$ (in unit of erg cm$^{-2}$
s$^{-1} $sr$^{-1}$) emitted  in the rest frame of the blob can be
boosted in the observer frame by a following scaling,
\begin{eqnarray}
    \label{eq:boostednuFnu}
    \nu F_{\nu} (E_{\gamma}) \to \delta^{4} \nu F_{\nu} (E_{\gamma}/\delta),
\end{eqnarray}
where the  Doppler factor is represented by
\begin{equation}
\label{eq:delta}
\delta=\frac{1}{\Gamma (1-\beta \cos \theta)},
\end{equation}
which behaves like $\sim \Gamma$ for $\theta \simlt 1/\Gamma$,
 and
$\sim 1/\Gamma$ in the other limit for off-axis cases, $\theta \gg
1/\Gamma$. Indeed, this shift of energy for synchrotron emission can
fit the change of cutoff scale for the flare component.  In addition,
for inverse-Compton processes in the rest frame of the blob, the
target photon is boosted from   the observer frame, and its
distribution is modified to be
\begin{eqnarray}
    \label{eq:phtonDist}
    dE_{\gamma} \frac{dn_{\gamma}}{dE_{\gamma}}(E_{\gamma} )     \to 
    \Gamma dE_{\gamma} \frac{dn_{\gamma}}{dE_{\gamma}}(E_{\gamma}/\Gamma).
\end{eqnarray}
Therefore  the inverse-Compton power is more enhanced than  the
synchrotron power by a factor of $\Gamma^{2}$  for the Thomson limit,
\begin{equation}
\label{eq:power}
\frac{(\nu F_{\nu})_{\rm IC}}{(\nu F_{\nu})_{\rm syn}}
\to \Gamma^2 \frac{(\nu F_{\nu})_{\rm IC}}{(\nu F_{\nu})_{\rm syn}}~~.
\end{equation}
Considering the Klein-Nishina effect, the enhancement could be smaller
than $\Gamma^2$.~\footnote{ Note that the magnetic field (inherent
in the blob) does not necessarily have the same boosting as the photon
field (penetrating into the blob).}

Next we discuss our choice of model parameters.  If  the maximum
energy of electron in the blob rest frame is limited by the
synchrotron cooling, the bulk Lorentz factor should be  $\simlt 10$
to explain   the energy shift of the maximum energy $E_{\rm
syn}^{\max}$    shown in equation~(\ref{eq:emax}) at the flare.
In that case, the
synchrotron cooling time should be shorter than the variability
timescale, which gives
\begin{equation}
\label{eq:gmax_cool}
E'_{{\rm max},e} > 170~{\rm TeV}~ \left( \frac{B'} {3~{\rm mG}} \right)^{-2}
\left( \frac{\delta} {10} \right)^{-1}
\left(\frac{\Delta t_{\rm obs}}{8~{\rm h}}\right)^{-1}~~,
\end{equation}
where $E'_{{\rm max},e}$ is  the maximum energy of electrons,
and $B'$ is magnetic field in the  rest frame of the blob,
respectively.   

The Lorentz factor can be larger than
$\simgt 10$ because the maximum energy of electrons is not
necessarily  limited only by cooling, but by the blob size (e.g.,
Ohira et al. 2011).  In this latter case, the Larmor radius of
electron with the maximum energy would be  comparable to the blob
size, $\sim \delta c \Delta t_{\rm obs}$ in the blob rest frame,
which gives
\begin{equation}
\label{eq:gmax_size}
E'_{{\rm max},e} =  790~{\rm TeV}~ \left( \frac{B'} {3~{\rm mG}} \right)
\left( \frac{\delta} {10} \right)
\left(\frac{\Delta t_{\rm obs}}{8~{\rm h}}\right)~~.
\end{equation}
Equating~(\ref{eq:gmax_cool}) with (\ref{eq:gmax_size}), we obtain a
threshold magnetic field  in the blob rest frame,
\begin{equation}
\label{eq:bc}
B_{\rm th}' =1.8~{\rm mG} \left(\frac{\delta}{10}\right)^{-2/3}
\left(\frac{\Delta t_{\rm obs}}{8~{\rm h}}\right)^{-2/3}~~.
\end{equation}
For $B'>B_{\rm th}'$, the maximum energy of electron is limited by the
synchrotron cooling.  However, $B_{\rm th}'$ is much larger than that
expected from the standard model of the steady-state Crab Nebula
(e.g. Kennel \& Coroniti 1984).  Therefore,  we also consider the
case, $B'<B_{\rm th}'$, where the maximum  energy of electron is
limited by the blob size.   Because the observed energy of synchrotron
photons for a monoenergetic electron is written as 
\begin{eqnarray}
    \label{eq:Esyn}
    E_{\rm syn} = 95~{\rm MeV} 
    \left( \frac{\delta} {10} \right)
    \left( \frac{E'_{{\rm max},e}} {500~{\rm TeV}} \right)^{2}
    \left( \frac{B'} {3~{\rm mG}} \right),
\end{eqnarray}
by removing $E'_{{\rm max},e}$ from
Equation~(\ref{eq:gmax_size}),  the magnetic field in the blob rest
frame is obtained  as
\begin{eqnarray}
    \label{eq:Blow2}
        B' = 2.2~{\rm mG} 
    \left( \frac{E_{\rm syn}} {10^2~{\rm MeV}} \right)^{1/3}
    \left( \frac{\delta}{10} \right)^{-1}
    \left( \frac{\Delta t_{\rm obs}}{8~{\rm h}} \right)^{-2/3}.
\end{eqnarray}
Then the  condition  $B' < B_{\rm th}'$ gives lower bound on the
Doppler factor
\begin{equation}
\label{eq:size}
\delta > 22~\left( \frac{E_{\rm syn}} {10^2~{\rm MeV}} \right).
\end{equation}
%
By removing $\delta$ in (\ref{eq:gmax_size})  with
(\ref{eq:Esyn}), we  get the maximum energy of electron in
the blob rest frame,
\begin{eqnarray}
    \label{eq:gammaUpp}
E'_{{\rm max},e} =  480~{\rm TeV}
    \left( \frac{E_{\rm syn}}{10^{2}~{\rm MeV}} \right)^{1/2}
    \left( \frac{\Delta t_{\rm obs}}{8~{\rm hours}} \right)^{1/3},
\end{eqnarray}
 which depends only on observables.
In  this size-limit case, the variability timescale should  be
determined by the dynamics of the blob, since the cooling timescale is
longer than that.  This may be favorable for explaining  the
comparable timescales of rise and decay in the observed flares.

In Fig~\ref{fig:f5} we plot  theoretical calculations of the spectrum
fitted to the flare component, adopting a scaling law for magnetic
field $B' = 220 \mu{\rm G} (\delta/10^{2})^{-1}$ in  equation
(\ref{eq:Blow2}) and the maximum energy of electron  obtained in
equation (\ref{eq:gammaUpp}).  The thick solid line  shows a result
with $\Gamma = \delta = 10^{2}$ and $B=223~\mu$G.  We call this set of
parameters ``fiducial model''. Because the  electrons in the blob
cannot  be cooled  in such a short time, the cooling break does not
appear in GeV region. Then the spectral index of electron  could be
the same as the initial value, $n_{s} \sim $2.35.  Consequently the
photon index of synchrotron radiation  ($F_{\nu}/\nu$) at around
${\cal O}(10^{2})$~MeV becomes $(n_{s}+1)/2 \sim 1.6$, which is
consistent with the observation ($1.27 \pm 0.12$) reported by
Buehler~(2011).   Below ${\cal O}$(10)~MeV,  the synchrotron flare
component is smaller than the stationary one.  We also plot  the high
and low $\Gamma$ models  with $\Gamma=\delta=700$ and
$\Gamma=\delta=30$ by using the same scaling law for the magnetic
field.

On the other hand, in TeV-PeV region, it is remarkable that the
inverse-Compton radiation of the flare component can exceed the
stationary one significantly at $E_{\gamma} \simgt {\cal O}(10)$~TeV
for the fiducial model.  That is also due in part to that the
inverse-Compton power is enhanced more than the synchrotron power
because of the boosted target photon distribution by a factor of
$\Gamma^{2}$ shown in (\ref{eq:power}).~\footnote{When we use the
scaling $B'\propto \delta^{-1}$ in equation (\ref{eq:Blow2}) in order
to fit the synchrotron component in GeV region, the magnetic energy
density is reduced by $\delta^{-2}$.  With (\ref{eq:power}), the
inverse-Compton component is enhanced by a factor of $\Gamma^{2}
\delta^{2}$ in total.}.  From Figure \ref{fig:f5}, we find that higher
Lorentz factor than $\Gamma \simgt 700$ has been already excluded by
the current GeV observations and $\Gamma \simgt 200$ by TeV
observations (Mariotti et al. 2010; Ong et al. 2010).\footnote{We
have obtained this upper bound on the Lorentz factor by using the
observational upper bounds on the flux at GeV and TeV with our
estimates of the inverse Compton emission.}

The future gamma-ray observatories such as CTA,  Tibet AS + MD  and LHAASO will
be able to probe $\Gamma$ down to $\sim 30$ by 8 hours observations
(by one night).  The sensitivities are shown in  Fig.~\ref{fig:f5} by
conservatively linearly scaling them from 50 hours to 8 hours.  
 Even if the flare flux is less than the stationary one,
we can detect it down to the statistical error of photons.
Note
that the ratio of inverse Compton to synchrotron power depends only on
$\Gamma$ in equation (\ref{eq:power}), not on $\delta$, so that we can
constraint $\Gamma$ even for an off-axis event.  Note also that  the
cooling effects can be seen  a little in $\simgt$ PeV region  for
$\Gamma \simlt$ 30.

To be more conservative, we also  show similar calculations with adding
an artificial low-energy cutoff for  the electron spectrum where we took
zero flux for $E_{e}<E_{e,{\rm cut}}=10^{2}$~TeV  in the blob rest
frame, which is represented by the shallower thick solid lines  [this
may explain the X-ray feature (Tavani et al. 2011)].
Even in this case, we can probe $\Gamma$ down to $\sim 700$.
 
Therefore we can discern the model from  the nonrelativistic ones for
the flares at ${\cal O}$(1) GeV by observing the inverse-Compton
radiation in the ${\cal O}$(10)TeV--${\cal O}$(1)PeV energies.

Here it should be meaningful to check an energy ratio of electron
to magnetic field in the blob.
The flare luminosity in the
blob rest frame, $L'$, is  given
 by
\begin{equation}
\label{eq:luminosity1}
L'=4\pi d^2 (\nu F_{\nu})_{\rm obs} \delta^{-4}~~.
\end{equation}
Assuming that this luminosity is  produced by synchrotron radiation of
electrons with  a typical maximum energy, 
the luminosity is written as
\begin{equation}
\label{eq:luminosity2}
L'=N_{\rm e}'(E'_{{\rm max},e}) \times \frac{4}{3}\sigma_{\rm T} c \left(\frac{B'^2}{8\pi}\right) 
\left(\frac{E_{\max,e}'}{m_e c^2}\right)^2 ~~,
\end{equation}
where $N_{\rm e}'(E'_{{\rm max},e})$ is the number of electrons with
the maximum energy in the blob rest frame.
From Equations~(\ref{eq:Blow2}), (\ref{eq:gammaUpp}), (\ref{eq:luminosity1}) and (\ref{eq:luminosity2}), we obtain the total energy of electrons
with $E_{\max,e}'$,
%
\begin{eqnarray}
U_{e}'&=&N_{e}'(E'_{{\rm max},e})
E'_{\max,e}
\\
   &=& 1.3 \times 10^{37}~{\rm erg}~\left(\frac{E_{\rm syn}}{10^2~{\rm MeV}}\right)^{-7/6} \left(\frac{\delta}{10}\right)^{-2}  \left(\frac{\Delta t_{\rm obs}}{8~{\rm h}}\right). \nonumber
\end{eqnarray}
On the other hand, from equation~(\ref{eq:Blow2}), the total energy of
the magnetic field in the blob rest frame is given by
\begin{eqnarray}
U_{B}'&=&\frac{B'^2}{8\pi} \times \frac{4}{3}\pi \left(\delta c\Delta t_{\rm obs}\right)^{3} \\
   &=& 4.8 \times 10^{41}~{\rm erg}~\left(\frac{E_{\rm syn}}{10^2~{\rm MeV}}\right)^{2/3} \left(\frac{\delta}{10}\right) \left(\frac{\Delta t_{\rm obs}}{8~{\rm h}}\right)^{5/3}.
\end{eqnarray}
 for the size-limit case $B'<B'_{\rm th}$.
Then, we find the ratio of the electron energy to the magnetic energy
to be
\begin{eqnarray}
\frac{U_{e}'}{U_{B}'}= 2.6 \times 10^{-5}~\left(\frac{E_{\rm syn}}{10^2~{\rm MeV}}\right)^{-11/6} \left(\frac{\delta}{10}\right)^{-3}  \left(\frac{\Delta t_{\rm obs}}{8~{\rm h}}\right)^{-2/3}. 
\label{eq:ratioU}
\end{eqnarray}
This ratio can increase by $\sim (E_{\max,e}/E_{\rm ib})^{n_s-2}$ for
$2<n_{s}<3$ if we include the low energy electrons.  Even in this
case, the boosted blob might be magnetically dominant.  Since the blob
position, $\Gamma \delta c\Delta t_{\rm obs} \sim 8.6\times 10^{14}
\Gamma \delta ~{\rm cm}$, may be also much smaller than the radius of
the termination shock, $\sim 3\times 10^{17}~{\rm cm}$, the blob might
be produced in the pulsar wind before the magnetic energy is converted
to the bulk kinetic energy.  Alternatively, the observed flares might
be off-axis ones with small $\delta (\sim 1)$.  In this case we
predict larger flares than ever detected.  Or a flare may consist of
many pulses with $\Delta t_{\rm obs} < 8$ h, e.g., $U'_e\sim U'_B$ for
$\Delta t_{\rm obs} \sim 33$ msec (pulsar period).  It may be the
reason for the scarcity of flares that a flare needs many pulses (see
also Clausen-Brown \& Lyutikov 2012).

If we require that the blob energy $\Gamma U'_{B}$
is less than the spin-down energy during the flare 
${\dot E} \Delta t_{\rm obs} \delta/\Gamma$
in the observer frame,
we have
$\Gamma^2 \simlt 300 (E_{\rm syn}/10^2{\rm MeV})^{-2/3}
(\Delta t_{\rm obs}/8 {\rm h})^{-2/3}$
for the size-limit case $B'<B'_{\rm th}$.
Thus, if we find $\Gamma>30$,
we also imply that a flare consists of many pulses with 
$\Delta t_{\rm obs} < 8$ h.

We have not specified the radiation region, which could be the pulsar
wind or the shock at the knot of the inner nebula. We have inferred
the physical condition and found several possibilities, such as the
magnetically dominant case, the off-axis case, and the case of
superposition of many pulses, which may be discriminated by future
TeV-PeV observations as argued below Eq.~(\ref{eq:ratioU}).

As was mentioned in Introduction, so far there has been no consensus
of the theoretical models.  In the model with only increasing the
maximum energy of electron such as by the electric acceleration, there
is surely an excess in  PeV region for inverse-Compton flare
component. In order to detect this excess by  LHAASO, however, we
approximately need  a few tens of hours for the observation time,
which is longer than the typical duration of the flare.  In the model
with only increasing the magnetic field  such as by rapid compression,
the inverse-Compton flare is highly suppressed.


\section{Summary and Conclusion}

In order  to explain the origin of the  GeV flare in Crab Nebula, we
have studied  models in which a small blob is boosted, e.g.,  with a
Lorentz factor $\Gamma \simgt$~$30$, and emits synchrotron photon
higher than the maximum synchrotron energy $E_{\rm syn}^{\max}$ shown
in equation (\ref{eq:emax}).  We have also discussed possibilities
that   we will discern the model from the others such as
nonrelativistic models, by observing the corresponding inverse-Compton
flare component.  We have pointed out that  the inverse-Compton flare
can appear  in $\simgt {\cal O}$(1)~TeV region accompanied with  the
GeV flare in this kind of the boosted blob models with large Lorentz
factor because  the inverse-Compton power is more boosted than the
synchrotron power by $\sim \Gamma^{2}$.  High $\Gamma$ models have
been already  excluded  for $\Gamma \simgt 200$ by the current TeV
observations  and will be further down to $\Gamma \sim 30$ by the
future TeV--PeV observatories, such as CTA, Tibet AS + MD   or LHAASO.
 In addition, by considering this enhancement in the TeV-PeV region, in
near future we may  observe ``orphan TeV-flares'', which do not have
even a GeV flare.

Even  for the stationary component of Crab Nebula, we have also
pointed out for the first time that 
the absorption of PeV
photons by CMB radiation through pair creation $\gamma+\gamma_{\rm
CMB} \to e^{+} + e^{-}$
is important.
We must
consider this  effect whenever we fit the spectrum of Crab Nebula in
the ${\cal O}$(1)PeV regions.

It is notable that we will be able to accomplish those studies for
observation of Crab Nebula  at ${\cal O}$(1)TeV--${\cal O}$(1)PeV
energies  by using the future  gamma-ray telescopes such as CTA,
Tibet AS + MD  or LHAASO. We hope the earliest possible completions of
this kind of new  gamma-ray telescopes.

\section*{Acknowledgments}
We thank F. Takahara, and S. Tanaka for useful
discussions.  This work is supported in part by grant-in-aid from the
Ministry of Education, Culture, Sports, Science, and Technology (MEXT)
of Japan, No. 21111006, No. 23540327 (K.K.), No.22244030 (K.I. and
K.K.), No.21684014 (K.I. and Y.O.),  No. 22244019 (K.I.).
K.K. was partly supported by the
Center for the Promotion of Integrated Sciences (CPIS) of Sokendai,
No.~1HB5806020.


\bsp
\label{lastpage}
\end{document}